\documentclass[amsmath, amsfonts, superscriptaddress, showpacs, twocolumn, prb]{revtex4}
\usepackage{graphicx}
\usepackage{epsfig}
\usepackage{bm}
\usepackage{dcolumn}
\usepackage{amsmath}
\usepackage{amssymb}
\usepackage{color}

\begin{document}

\title{Quantitative analysis of quantum phase slips in superconducting Mo$_{76}$Ge$_{24}$
nanowires\\
revealed by switching-current statistics}

\author{T.~Aref}
\affiliation{Department of Physics, University of Illinois at
Urbana-Champaign, Urbana, Illinois 61801, USA}

\affiliation{Low Temperature Laboratory, Aalto University, 00076
Aalto, Finland}

\author{A.~Levchenko}
\affiliation{Department of Physics and Astronomy, Michigan State
University, East Lansing, Michigan 48824, USA}

\author{V.~Vakaryuk}
\affiliation{Materials Science Division, Argonne National
Laboratory, Argonne, Illinois 60439, USA}

\author{A.~Bezryadin}
\affiliation{Department of Physics, University of Illinois at
Urbana-Champaign, Urbana, Illinois 61801, USA}

\begin{abstract}
We measure quantum and thermal phase-slip rates using the standard
deviation of the switching current in superconducting nanowires. Our
rigorous quantitative analysis provides firm evidence for the
presence of quantum phase slips (QPSs) in homogeneous nanowires at
high bias currents. We observe that as temperature is lowered,
thermal fluctuations freeze at a characteristic crossover
temperature $T_q$, below which the dispersion of the switching
current saturates to a constant value, indicating the presence of
QPSs. The scaling of the crossover temperature $T_q$ with the
critical temperature $T_c$ is linear, $T_q\propto T_c$, which is
consistent with the theory of macroscopic quantum tunneling. We can
convert the wires from the initial amorphous phase to a
single-crystal phase, \textit{in situ}, by applying calibrated
voltage pulses. This technique allows us to probe directly the
effects of the wire resistance, critical temperature, and morphology
on thermal and quantum phase slips.
\end{abstract}

\date{June 26, 2012}


\maketitle

\section{Introduction}

Topological fluctuations of the order parameter field, so-called
Little's phase slips,~\cite{Little1} are at the heart of
superconductivity at the
nanoscale.~\cite{Tinkham,AGZ-Review,Bezryadin-Review} These
unavoidable stochastic events give rise to the finite resistivity of
nanowires below the mean-field transition temperature. Thermally
activated phase slips (TAPSs) have been routinely observed
experimentally; see Ref.~\onlinecite{Bezryadin-Review} for a review.
However, at low temperatures, phase-slip events are triggered by
intrinsic quantum
fluctuations,~\cite{MDC-PRB87,Giordano,Sahu-NatPhys09} so they are
called quantum phase slips (QPSs) and represent a particular case of
macroscopic quantum tunneling (MQT). A clear and unambiguous
demonstration of MQT in homogeneous superconductors is of great
importance, from both the fundamental and the technological
prospectives. It has been argued recently by Mooij and
Nazarov~\cite{MN-NatPhys06} that a wire where coherent QPSs take
place may be regarded as a new circuit element, the phase-slip
junction, which is a dual counterpart of the Josephson
junction.~\cite{Pop-NatPhys10} The proposed phase-slip
qubit~\cite{MH-NJP05} and other coherent
devices~\cite{MN-NatPhys06,HN-PRL11,Zorin-PRL12,Astafiev-NP12} may
be useful in the realization of a new current standard. Furthermore,
comprehensive study of QPSs may elucidate the microscopic nature of
superconductor-insulator quantum phase transition in
nanowires.~\cite{Dynes,BezryadinNature,Shahar,Bollinger1}

It is difficult to obtain firm conclusions about the presence of
QPSs by means of low-bias resistance measurements because the
resistance drops to zero at relatively high temperatures. Measured
in the linear transport regime, high-resistance wires seemingly
exhibit QPSs,~\cite{Lau} while low-resistance wires probably do
not.~\cite{Bollinger2} For high-bias currents, on the other hand,
Sahu \textit{et al.}~\cite{Sahu-NatPhys09} obtained strong evidence
supporting the quantum nature of phase slips by measuring
switching-current distributions. The observed drop of the
switching-current dispersion with increasing temperature was
explained by a delicate interplay between quantum and multiple
thermal phase slips. Recently Li \textit{et al.}~\cite{Li-PRL11}
provided direct experimental evidence that, at sufficiently low
temperatures, {\it each} phase slip causes nanowire switching from
superconducting to the normal state by creating a hot
spot.~\cite{Sahu-NatPhys09,Shah-PRL07} The destruction of
superconductivity occurs by means of overheating the wire caused by
a single phase slip. Thus the dispersion of phase-slip events is
equivalent to the dispersion of the switching current.

We build on these previous findings and reveal MQT in homogeneous
nanowires via the quantitative study of current-voltage
characteristics. First, we examine the higher-temperature regime,
$T_q<T<T_c$, and identify thermal phase slips through the
temperature dependence of the switching-current standard deviation
$\sigma$, which obeys the 2/3 power law predicted by
Kurkij\"{a}rvi.~\cite{Kurkijarvi} At lower temperatures, $T<T_q$, a
clear saturation of $\sigma$ is observed; this behavior is
indicative of MQT. Important evidence in favor of QPSs is provided
by the fact that the mean value of the switching current keeps
increasing with cooling even when the associated dispersion is
already saturated. We observe a linear scaling of the saturation
temperature $T_q$ with the critical temperature $T_c$ of the wire.
We also show that such behavior is in agreement with our
generalization of the MQT theory. This fact provides extra assurance
that other mechanisms, such as electromagnetic (EM) noise or
inhomogeneities, are not responsible for the observed behavior.
Furthermore, we achieve \textit{controllable tunability} of the wire
morphology by utilizing a recently developed voltage pulsation
technique.~\cite{Aref-NanoTech11} The pulsation allows us to
gradually crystallize the wire and to change its $T_c$ \textit{in
situ}.  The fact that the QPS manifestations are qualitatively the
same in both amorphous and crystallized wires eliminates the
possibility that the observed MQT behavior is caused by the presence
of weak links. Thus we provide conclusive evidence for the existence
of QPSs in homogeneous wires in the nonlinear regime of high-bias
currents.

\section{Experimental details}

Superconducting nanowires were fabricated by molecular
templating.~\cite{BezryadinNature,Bezryadin-Review} Briefly, a
single-wall carbon nanotube is suspended across a trench etched in a
silicon wafer. The nanotube and the entire surface of the chip are
then coated with 10-20~nm of the superconducting alloy
Mo$_{76}$Ge$_{24}$ using dc magnetron sputtering. Thus a nanowire,
seamlessly connected to thin film electrodes at its ends, forms on
the surface of the electrically insulating nanotube. The electrodes
approaching the wire are between $5$ and $20~\mu$m wide. The gap
between the electrodes, in which the nanowire is located, is 100~nm.

The signal lines in the He-3 cryostat were heavily filtered to
eliminate electromagnetic noise, using copper-powder and
silver-paste filters at low temperatures and $\pi$ filters at room
temperature.~\cite{MDC-PRB87} To measure switching-current
distributions, the bias current was gradually increased from zero to
a value that is about 20\% higher than the critical current
(1-10~$\mu$A). Such large sweeps ensure that each measured $I$-$V$
curve exhibits a jump from the zero-voltage state to the resistive
normal state. Such a jump is defined as the switching current
$I_{sw}$, and $N=10^{4}$ switching events were detected at each
temperature through repetitions of the $I$-$V$ curve measurements
$N$ times. The standard deviation (i.e., dispersion) $\sigma$ and
the mean value $\langle I_{sw}\rangle$ are computed in the standard
way.

\begin{figure}
 \includegraphics[width=8cm]{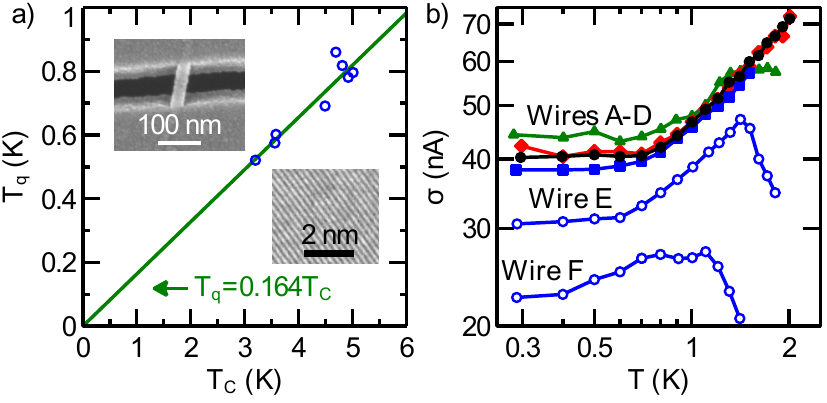}\vskip-.45cm
  \caption{[Color online] (a) The saturation temperature
  $T_q$ vs the critical temperature $T_c$ for samples
  A-D, pulsed and unpulsed. The line is the best fit. The top insert
  shows SEM image of an unpulsed nanowire; the bottom insert shows
  a TEM micrograph of a nanowire crystallized by applying
  voltage pulses.~\cite{Aref-NanoTech11} The fringes corresponding to atomic planes are visible.
  (b) The standard deviation of the switching current versus temperature,
  for samples A-F (prior to any pulsing).
  }
  \label{Fig2}\vskip-.5cm
\end{figure}

We apply strong voltage pulses to induce Joule heating, which
crystallizes our wires [see bottom inset in Fig.~\ref{Fig2}(a)] and
also changes their critical temperature
$T_c$.~\cite{Aref-NanoTech11} With increasing pulse amplitude, $T_c$
(as well as $I_c$) initially diminishes and then increases back to
the starting value or even exceeds it in some cases. Such
modifications of $T_c$ and $I_c$ have been explained by
morphological changes, as the amorphous molybdenum germanium
(Mo$_{76}$Ge$_{24}$) gradually transforms into single-crystal
Mo$_3$Ge, caused by the Joule heating brought about by the voltage
pulses. The return of $T_c$ and $I_c$ is accompanied by a drop in
the normal resistance $R_n$ of the wire, which is caused by the
crystallization and the corresponding increase of the electronic
mean free path. The pulsing procedure allows us to study the effect
of $T_c$ on $T_q$ [see Fig.~\ref{Fig2}(a)] and the effect of the
morphology of the wire on the QPS process in general. Note that
after the pulsing is done and the morphology of the wire is changed
in the desired way, we always allow a sufficient time for the wire
to return to the base temperature before measuring $I_{sw}$.

\section{Results, Analysis and Modeling}

\begin{figure}
  \includegraphics[width=7cm]{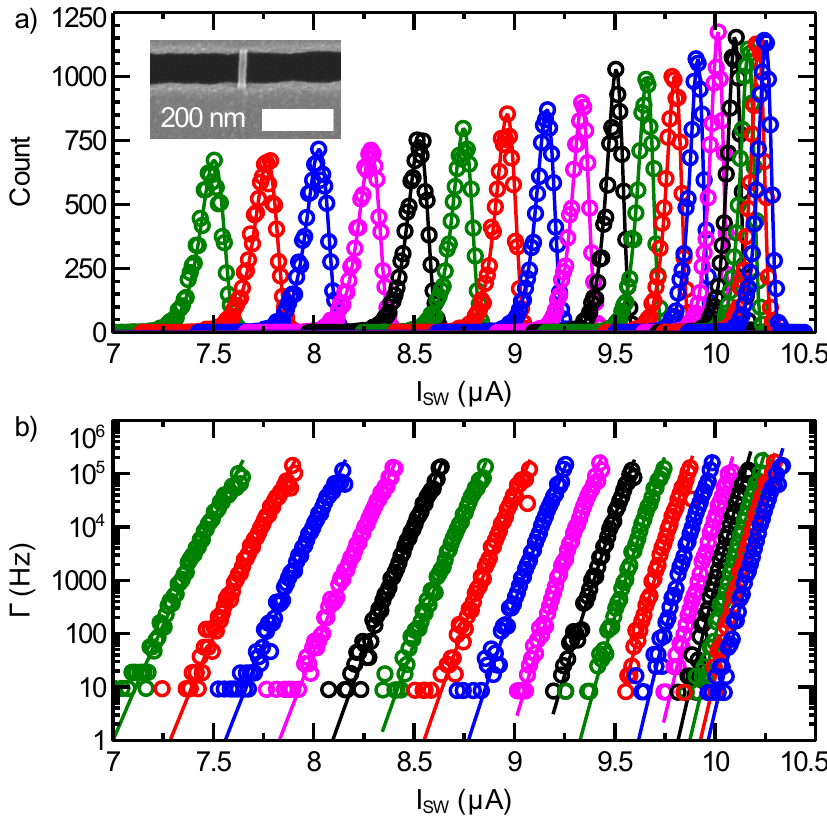}\vskip-.45cm
  \caption{[Color online] Distributions and the
  switching rates for wire A. (a) Measured switching
  current distributions (circles) for various temperatures
  ranging from 2~K for the left curve to 0.3~K for the
  right curve (step = 0.1~K). The fits are shown
  as solid lines of the same color.~\cite{Gumbel}
  The inset shows a SEM image of a representative nanowire after completing the pulsing procedure.
  (b) Switching rates, derived from the distribution shown in (a), are represented by
  circles, while solid curves of the same color are fits by
  Eq.~\eqref{gamma} with $b=3/2$.
}\label{Fig3}\vskip-.5cm
\end{figure}

Current-voltage characteristics for our wires display clear
hysteresis, sustained by Joule heating, similar to
Refs.~\onlinecite{Sahu-NatPhys09,Bollinger1,Tinkham-PRB03}. The
switching current from dissipationless branch to resistive branch of
the $I$-$V$ curve fluctuates from one measurement to the next, even
if the sample and the environment are unchanged. Examples of the
distributions of the switching current are shown in
Fig.~\ref{Fig3}(a) for different temperatures. Since, by definition,
the area under each distribution is constant, the fact that at
$T<0.7\,$K its height stops increasing with cooling implies that its
width, which is proportional to $\sigma$, is constant as well; see
Fig.~\ref{Fig2}(b). Thus we get the first indication that the
quantum regime exists for $T<0.7\,$K, i.e.~for this case
$T_q\approx0.7\,$K.

\begin{figure}[t!]
 \includegraphics[width=7.5cm]{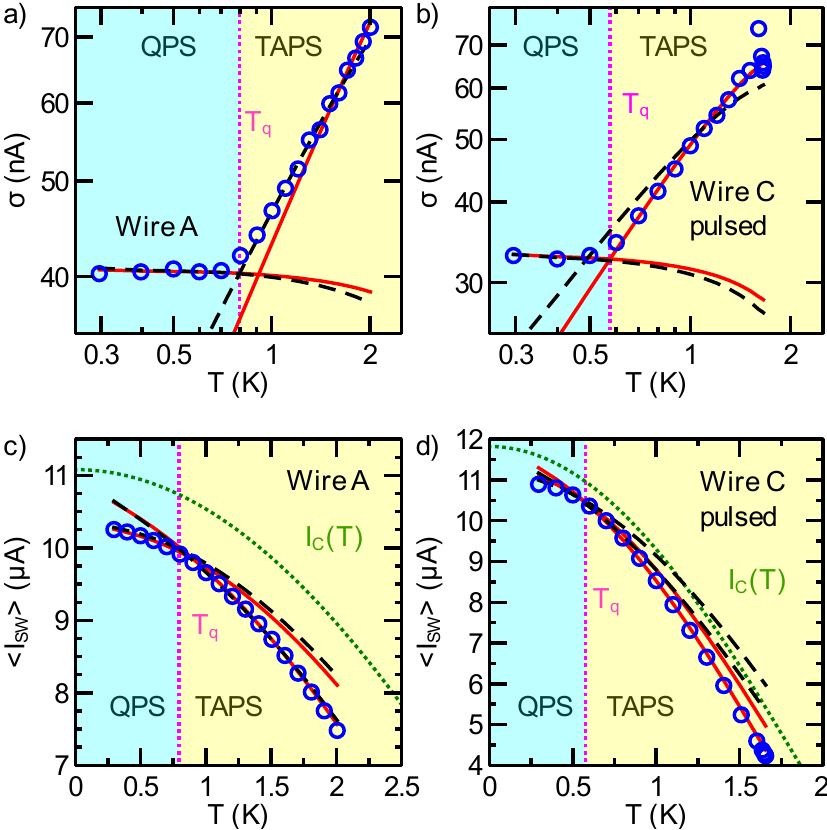}\vskip-.45cm
  \caption{[Color online] (c) and (d) The average switching current and (a) and (b) its standard
  deviation are plotted vs temperature. The computed
  critical current $I_c(T)$ is also plotted for comparison in (c) and (d).
  (a) Sample A, unpulsed. (b) Sample C, pulsed.
  In (a) and (b) the fits are generated
  by Eq.~\eqref{sigma}. The two almost-horizontal curves
  (solid and dashed), fitting well the low-temperature
  part, correspond to the QPS-dominated regime. They are computed assuming
  $T_{esc}=T_q$ in Eq.~\eqref{sigma}, where $T_q=$0.8~K for sample A and $T_q=$0.6~K
  for sample C. The two other curves (solid and dashed),
  which fit well the high temperature part of the data,
  represent
  TAPS according to Eq.~\eqref{sigma}, with $T_{esc}=T$.
  The solid red curve corresponds to $b=5/4$, and
  the dashed black curve corresponds to $b=3/2$.
  (c) Unpulsed $\langle I_{sw}\rangle$ and (d) pulsed $\langle I_{sw}$.
  $T_q$ is indicated by the vertical dotted line.
  The fits to $\langle I_{sw}\rangle$ are also shown, following the
  convention explained in (a) and (b), according to Eq.~\eqref{Isw}. The green dotted
  line is $I_c(T)$ from Bardeen's expression, which is
  used to compute $\langle I_{sw}\rangle$. Note that
  $\langle I_{sw}\rangle$ {\it does not} saturates at
  $T_q$ and keeps increasing for lower $T$.}\label{Fig1}\vskip-.5cm
\end{figure}

We now turn to the discussion and analysis of the main results.
Following the Kurkij\"arvi-Garg (KG)
theory~\cite{Kurkijarvi,Garg-PRB95} the rate of phase
slips,~\cite{FD} such as shown in Fig.~\ref{Fig3}(b), can be written
in the general form
\begin{equation}
    \label{gamma}
    \Gamma=\Omega\exp[-u(1-I/I_c)^b]\,,
\end{equation}
where $I$ and $I_c$ are the bias and critical currents,
respectively, $\Omega=\Omega_0 (1-I/I_c)^a$ is the attempt
frequency, and $u=U_c(T)/T_{esc}$, where $U_c$ is a model-dependent
free-energy barrier for a phase slip at $I=0$. Parameter $T_{esc}$
is known as the effective escape temperature. In the case of thermal
escape, $T_{esc}=T$, according to Arrhenius law, where $T$ is the
bath temperature. In the quantum fluctuation-dominated regime
$T_{esc}$ is the energy of zero-point fluctuations. We have checked
explicitly that this energy equals the crossover temperature $T_q$
(see the Appendix for details). Thus in the QPS regime
$T_{esc}=T_q$.

Exponent $b$ defines the dependence of the phase-slip barrier on
$I$. While the value of this exponent is well known for thermally
activated phase slips, in the quantum regime the value of $b$ is
poorly understood. Thus experimental determination of $b$ represents
a significant interest to the community. The approximate linearity
of the semi logarithmic plots $\Gamma (I)$ (see the Appendix for
details), which is especially pronounced at low temperatures in the
QPS regime [curves on the right in Fig.~\ref{Fig3}(b)], provides a
useful estimate for the current exponent $b_\textsc{qps}\sim 1$.

As was shown in Refs.~\onlinecite{Sahu-NatPhys09,Li-PRL11} and
\onlinecite{Shah-PRL07}, a single phase-slip event is sufficient to
drive a nanowire into the resistive state so that the temperature
dependence of the dispersion is power law. In all our
high-critical-current samples (unpulsed samples A--D, and also
C-pulsed, and D-pulsed) the power law is observed, as is illustrated
in Fig.~\ref{Fig1} for two representative samples (see the range
$T_q<T<$2~K).

As the temperature is lowered, the TAPS rate drops exponentially,
while the QPS rate remains finite. This leads to the crossover
between thermal and quantum regimes, which occurs at $T_q$. It will
be shown below that a definite relation exists between the
superconducting transition temperature $T_c$ and $T_q$. We suggest
that experimental observation of such relation can be used as a tool
in identifying MQT. In particular, we use this approach to eliminate
the possibility of noise-induced switching and thus confirm the QPS
effect.

According to the KG theory~\cite{Kurkijarvi,Garg-PRB95}, the average
value of the switching current is given by
\begin{equation}
    \label{Isw}
    \langle I_{sw}\rangle\simeq I_c\left[1-u^{-1/b}\kappa^{1/b}\right]\,.
\end{equation}
Here $\kappa=\ln(\Omega_0t_\sigma)$, and $t_\sigma$ is the time
spent sweeping through the transition. Since $\Omega_0 t_\sigma$ is
present only in the logarithm, its exact value is fairly
unimportant. Dispersion $\sigma$ of the switching current which
corresponds to the escape rate in Eq.~(\ref{gamma}) can be
approximated as
\begin{equation}
\label{sigma} \sigma\simeq\frac{\pi
I_c}{\sqrt{6}b}u^{-1/b}\kappa^{(1-b)/b}=\frac{\pi
I_c}{\sqrt{6}b\kappa}\left[1-\frac{\langle
I_{sw}\rangle}{I_c}\right]\,.
\end{equation}

Let us discuss first the higher-temperature TAPS regime. To
distinguish the Josephson junction (JJ) from the phase-slip junction
(PSJ), as we call our superconducting nanowire following
Ref.~\onlinecite{MN-NatPhys06}, we consider in parallel two basic
models. The JJs are commonly described by the McCumber-Stewart
model~\cite{McCumberStewart,FD} with the corresponding washboard
potential. It can be solved exactly and gives $U_c=2\sqrt{2}\hbar
I_c/3e$ and $b=3/2$. The PSJ barrier for the current-biased
condition,~\cite{Tinkham-PRB03,McCumberGibbsenergy} which is our
case, is $U_c=\sqrt{6}\hbar I_c/2e$ and the power is $b=5/4$.
Although $U_c$ is very close in both models, it is expected that
different scaling determined by $b$ will translate into different
current switching dispersions.

Figures.~\ref{Fig1}(a) and \ref{Fig1}(b) show our main results for
the temperature dependence of the standard deviation for one
representative unpulsed wire and one pulsed wire (see the Appendix
for more information). In all the cases $\sigma(T)$ decreases as a
power law and saturates to a constant value at low temperatures. The
higher-temperature regime of TAPS appears in good agreement with the
KG theory. All our amorphous wires show properties somewhat similar
to JJs ($b_{\mathrm{TAPS}}=3/2$), indicating that the barrier for
phase slips depends on the bias current as $(1-I/I_c)^{3/2}$. The
two pulsed and crystallized wires agree better with the predictions
of PSJ model for perfectly homogeneous one-dimensional (1D) wires
($b_{\mathrm{TAPS}}=5/4$). The QPS phenomenon is present in both
types of wires, as is evidenced by the observed saturation of the
dispersion. Thus we conclude that the QPS is ubiquitous, as it
occurs in amorphous wires and in 1D crystalline wires. Note that the
pulsed crystalline wires are more into the 1D limit since their
coherence length is larger while their diameter, measured under
scanning electron microscopy (SEM), is not noticeably affected by
the pulsing crystallization [see inset in Fig.~\ref{Fig3}(a)].

Now let us focus on the quantum fluctuations represented by the
saturation of $\sigma$ at low temperatures $T<T_q$. The observed
crossover is a key signature of MQT. Strong evidence that the
saturation is not due to any sort of EM noise or an uncontrolled
overheating of electrons above the bath temperature follows from the
fact that although $\sigma$ is constant at $T<T_q$, the switching
current keeps growing with cooling, even at $T<T_q$ [see
Figs.~\ref{Fig1}(c) and \ref{Fig1}(d)]. The observed saturation of
$\sigma$ for $T<T_q$ and the simultaneous increase of $\langle
I_{sw} \rangle$ with cooling at $T<T_q$ are in agreement with the
QPS theoretical fits of the KG theory (Fig.~\ref{Fig1}). The value
of the critical current here is taken from Bardeen's
formula~\cite{Bardeen-RMP62}: $I_c=I_{c0}(1-(T/T_c)^2)^{3/2}$, which
works well at all temperatures below $T_c$.~\cite{Brenner-PRB11} The
critical current at zero temperature $I_{c0}$ and $T_c$ are used as
fitting parameters. Such MQT-reassuring behavior (i.e., saturation
of $\sigma$ when $\langle I_{sw}\rangle$ does not show saturation)
has not been observed previously on superconducting nanowires and
constitutes our key evidence for QPSs.

Conventionally, the crossover temperature $T_q$ between regimes
dominated by thermal or quantum phase slips is defined as a
temperature at which the thermal activation exponent becomes equal
to the quantum action, both evaluated at zero-bias current.
\cite{GZTAPS,Khlebnikov} Such definition is limited to small-bias
currents; thus it is not applicable to our study since it neglects
the role of the bias current, which in our case is the key control
parameter.\cite{Note,Khlebnikov-ArXiv}

Alternatively, the strength of a phase-slip mechanism can be
described by the deviation of the average switching current from the
idealized critical current of the device $I_c$, which is the
switching current in the absence of stochastically induced phase
slips. Such a characterization provides an assessment of the
tunneling rate since it is the latter which determines $\langle
I_{sw} \rangle$. Using $I_c - \langle I_{sw} \rangle$ as a measure
of a phase-slip tunneling rate and accounting for the fact that the
idealized critical current of the device is a phase-slip-independent
quantity, we arrive at the following implicit definition of the
crossover temperature $T_q$: $\langle I_{sw,1} (T_q) \rangle=\langle
I_{sw,2} (T_q) \rangle$ where 1 and 2 denote two phase slip-driving
mechanisms. Assuming that $\langle I_{sw,i} \rangle $  can be
represented by a generic expression (\ref{Isw}) and that parameters
$\Omega_0$, $a$, $u$, and $b$ can be specified for a particular
phase-slip mechanism, the above equation reduces to
\begin{equation}
    u_1^{1/b_1}(T_q) = \gamma \,    u_2^{1/b_2}(T_q).
    \label{B=B}
\end{equation}
Constant $\gamma \equiv \kappa_1^{1/b_1}  / \kappa_2^{1/b_2}$
depends only logarithmically on temperature and other parameters;
such dependence is subleading and will be neglected. \footnote{It
can be shown that the definition of $T_q$ through tunneling rates
$\Gamma_i$ as described above also leads to Eq.~(\ref{B=B}) with,
however, different value of $\gamma$.}

To calculate $T_q$ using Eq.~(\ref{B=B}) knowledge of phase-slip
parameters $u_i$ and $b_i$ is required. For a long wire in the TAPS
regime  these are given by  $u_{\textsc{taps}} = (11.34/T) s N_0
\sqrt{D} (T_c - T)^{3/2}$ and $b_{\textsc{taps}}=5/4$, where $s$ is
the wire cross section, $D$ is the diffusion coefficient, and $N_0$
is the density of states.\cite{GZTAPS} In the QPS regime
$u_{\textsc{qps}} = A \, s N_0 \sqrt{D \Delta}$ where $A$ is a
numerical constants of order 1 and $\Delta$ is the
temperature-dependent gap. \cite{GZTAPS,Khlebnikov} Since \textit{a
posteriori} $T_q \ll T_c$, one can safely approximate $\Delta$ by
its zero-temperature value $\Delta = 1.76 T_c$.

The value of $b_{\textsc{qps}}$  (the exponent which governs the
current dependence of the QPS action) is poorly known. Motivated by
the fact that the fits to rates $\Gamma$ shown on Fig.~\ref{Fig3}(b)
are made with the same value of $b$ for all temperatures and match
the data well, we make a plausible assumption that
$b_{\textsc{qps}}\approx b_{\textsc{taps}}$. Then, combining
Eq.~(\ref{B=B}) with the expressions for $u_{\textsc{qps}}$ and
$u_{\textsc{taps}}$ given above, one arrives at the conclusion that
$T_q \propto T_c$. This is in agreement with our experimental
finding that $T_q \approx 0.16 T_c$. The observed coefficient of
proportionality 0.16 implies that $\gamma^{b} A \approx 41$.
\footnote{It should be noted that the expression for
$u_{\text{QPS}}$ given by Golubev and Zaikin in
Ref.~\onlinecite{GZTAPS} is different by a numerical factor of order
5 from that used by Tinkham and Lau in
Ref.~\onlinecite{TinkhamLau-APL02}. Had we used the latter
expression the value of this product would be reduced by a
corresponding factor.}

In practice, when looking for MQT/QPSs through the temperature
dependence of the switching-current distribution, one has to worry
about the alternative explanation that the $\sigma$ saturation is
caused by the presence of a constant noise level. Such saturation,
if present, can also be analyzed in the framework outlined above.
Modeling noise as a thermal bath with temperature $T_n$ one obtains
that the crossover temperature to the noise-dominated phase-slip
regime is equal to $T_n$ and hence does not correlate with $T_c$,
which is in contrast to our observation, [Fig.~\ref{Fig2}(a)]. We
also argue that wires which are less susceptible to the noise, i.e.,
the wires with higher critical temperatures and therefore larger
barriers for phase slips, exhibit more pronounced quantum effects;
i.e., their saturation temperature $T_q$ is larger. We conclude
therefore that the correlation between the crossover temperature and
the critical temperature, observed in our experiment
[Fig.~\ref{Fig2}(a)], is strong evidence in favor of MQT below
$T_q$.

The saturation of $\sigma$ at low temperatures is seen on all tested
samples, A-F [Fig.~\ref{Fig2}(c)], which have critical currents of
11.1, 12.1, 13.1, 9.23, 5.9, and 4.3 $\mu$A, respectively (see
Appendix for additional data). Samples E and F have relatively low
critical currents. This fact leads to the occurrence of
multi-phase-slip switching events (MPSSE), manifested by the
characteristic drop of $\sigma$ with increasing $T$, observed at
higher temperatures. Such a drop was already observed on nanowires
with relatively low critical currents (between 1.1  and 6.1 $\mu$A)
in Refs.~\onlinecite{Sahu-NatPhys09,Li-PRL11}, which represents an
important consistency check for our findings. Here we focus on
samples with higher critical currents, which do not exhibit MPSSEs
and do not analyze our samples E and F, which exhibit MPSSEs
[Fig.~\ref{Fig2}(b)].

\section{Summary}

In summary, we demonstrate that in nanowires at moderately high
temperatures, $T>T_q$, the switching into the normal state at high
bias is governed by TAPSs. The corresponding standard deviation of
the switching current
 follows the Kurkij\"arvi-type power-law temperature
dependence $\sigma\propto T^{1/b}$. At low temperatures, $T<T_q$ the
dispersion of the switching distribution becomes temperature
independent. The crossover temperature $T_q$ from the TAPS- to the
QPS-dominated regime is proportional the wire's critical
temperature, in agreement with theoretical arguments. Thus QPSs are
unambiguously found in amorphous and single-crystal nanowires in the
regime of high bias currents, i.e., near the critical current.

\subsection*{Acknowledgment}
This material is based upon work supported by the DOE Grant No.
DEFG02-07ER46453 and by the NSF Grant No. DMR 10-05645. A.-L.
acknowledges support from Michigan State University.

\begin{figure}
  \includegraphics[width=8cm]{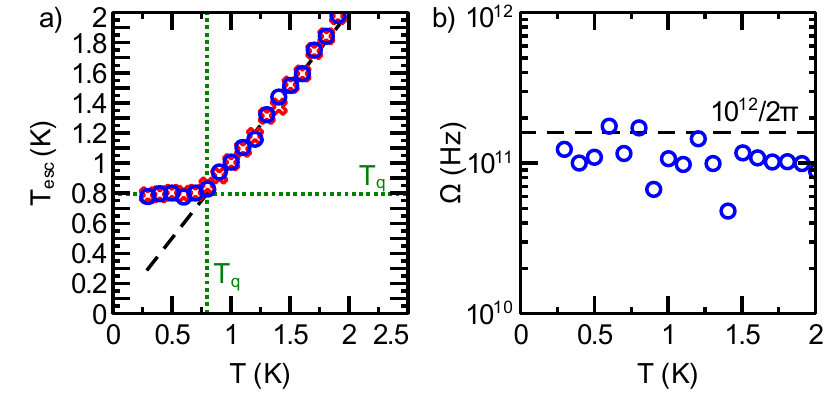}\vskip-.45cm
  \caption{[Color online] (a) The fitting parameter $T_{esc}$
  that defines the escape rate in Eq.~\eqref{gamma} presented as a function of
  temperature. (b) Temperature dependence of the escape frequency.
}\label{SOMFig1}\vskip-.5cm
\end{figure}

\section{Appendix}\label{Sec-App}

\textit{(a) Escape temperature and attempt frequency}. The fitting
parameter $T_{esc}$ for wire A is shown versus temperature in
Fig.~\ref{SOMFig1}a. For reference, the values of $T_q$, extracted
from the mean switching current and standard deviation fits, are
plotted on both horizontal and vertical scales as a dotted green
lines. One can clearly identify the regime of thermally dominated
escape $T_{esc}=T$ (shown by a black dashed line) above $T_q$ and
the regime of intrinsically quantum escape with an effective
temperature $T_{esc}=T_q$ at low temperatures.

\begin{figure}
  \includegraphics[width=5cm]{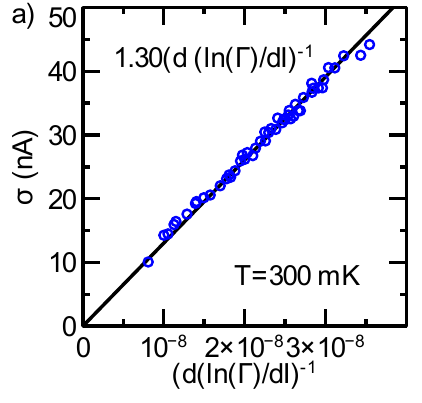}\vskip-.45cm
  \caption{[Color online] Standard deviation vs inverse
  $d\ln\Gamma/dI$ at base temperature $T=0.3$~K, which is already
  deep into the quantum regime.}\label{SOMFig2}
\end{figure}

Having measured $\sigma(T)$, one can invert Eq.~\eqref{sigma} to
find the corresponding $T_{esc}$ and perform the consistency check
for the theoretical model. The found $T_{esc}$ is plotted in
Fig.~\ref{SOMFig1}(a) as red crosses, which also matches well with
the escape temperature obtained by fitting the rates (shown as blue
circles).

In Fig.~\ref{SOMFig1}(b) we present the temperature dependence of
the attempt frequency introduced in Eq.~\eqref{gamma}. The dashed
line corresponds to the characteristic frequency
$2\pi\Omega=1/\sqrt{LC}\approx10^{12}$~s$^{-1}$, where
$L\approx1$~nH and $C\approx1$~fF are the kinetic inductance and
geometrical capacitance of the wire and the electrodes
correspondingly.

\begin{figure}
  \includegraphics[width=8cm]{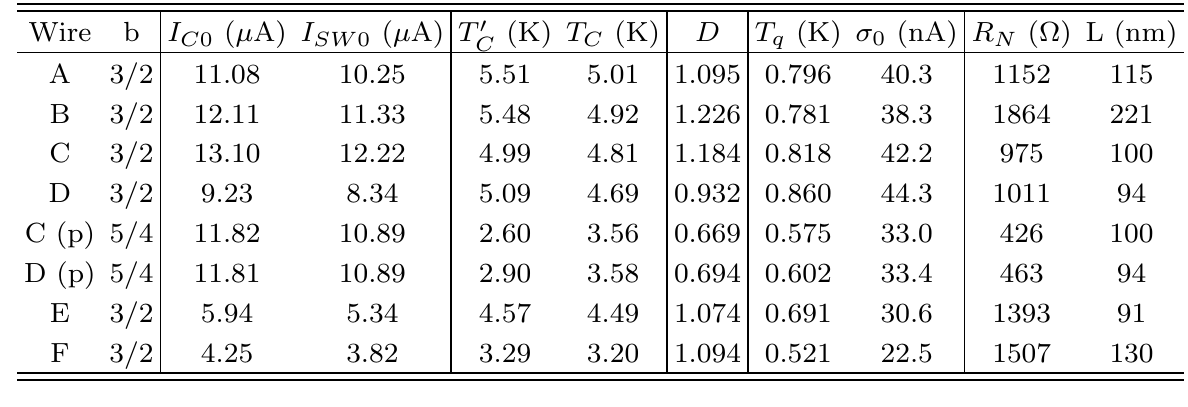}\vskip-.45cm
  \caption{Table of fitting parameters.}\label{Table}
\end{figure}

\textit{(b) Relation between $\sigma$ and $d\ln\Gamma/dI$}. We use
experimental data for the switching rates $\Gamma(I)$ from
Fig.~\ref{Fig3}b to check how the slope $d\ln\Gamma/dI$ relates to
the dispersion $\sigma$. Note that this slope is defined by the
slope of straight line fits in Fig.2(b) taken at the lowest
temperature. The results of such an analysis are presented in
Fig.~\ref{SOMFig2}. We find a linear dependence of the dispersion
with respect to the inverse slope of the semilogarithmic plots of
the switching rate versus the current. The result is in agreement
with the theorem proven in Ref.~\onlinecite{Bezryadin-Review}. The
best linear fit provides solid justification for the applicability
of the KG model in quantum regime, which we used for the
interpretation of our results.

\begin{figure}[h!]
  \includegraphics[width=7cm]{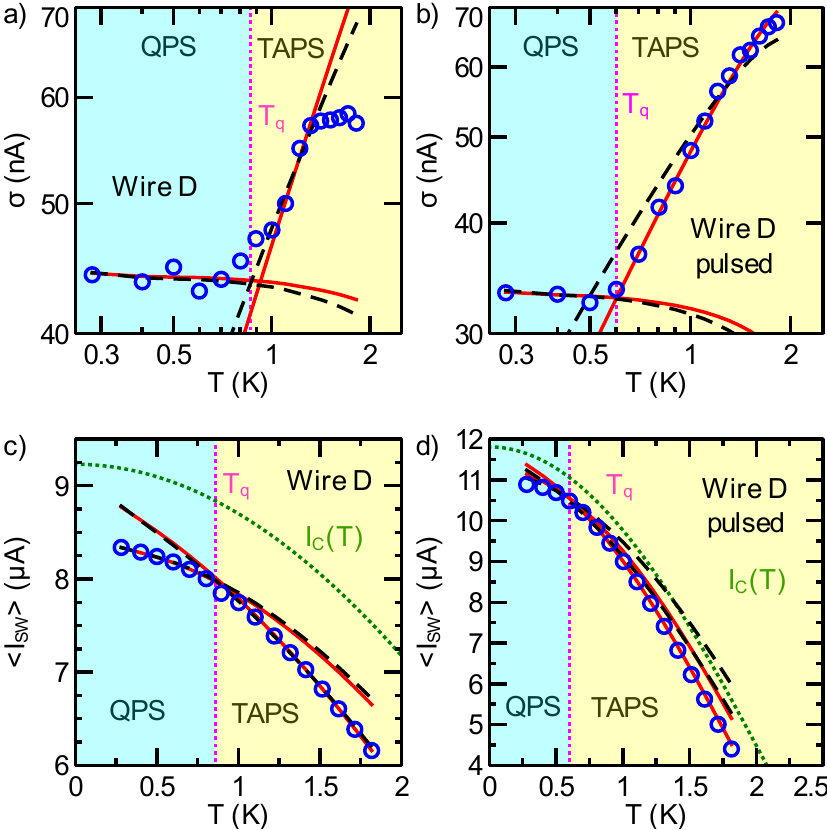}\vskip-.15cm
  \caption{[Color online] Standard deviations and critical currents versus temperature
  for wire D before and after pulsing. The convention for lines follow that explained
  in the caption of Fig.~\ref{Fig1} in the main text.}\label{SOMFig3}\vskip-.25cm
\end{figure}

\textit{(c) Fitting parameters}. Table shown in Fig.~\ref{Table}
summarizes all the fitting parameters used for the data analysis and
interpretation. The measurements were done for eight different wires
labeled from A to F. For wires C and D pulsation was applied, which
is indicated in Table by a subscript (p). The value of power
exponent $b$ which gave the best fit for the data is listed for
every wire. Note that for all wires the critical current at zero
temperature, $I_{c0}$, is slightly higher than that for the
switching current $I_{cw0}$ at base temperature. The critical
temperature used to fit the mean and standard deviation of the
switching current, $T'_c$, is relatively close to the critical
temperature used to fit the resistance versus temperature data.
$R(T)$ analysis was done by using result for TAPSs:
\begin{equation}\label{R(T)}
R(T)=R_n\exp[-\Delta F(T)/T],
\end{equation}
where $R_n$ is the normal state resistance of the nanowire, and
\begin{equation}\label{F(T)}
\Delta
F(T)=0.83\frac{R_q}{R_n}\frac{L}{\xi(0)}T_c[1-(T/T_c)^2]^{3/2}
\end{equation}
is the free-energy barrier for phase slips. Here $R_q=h/4e^2$ is the
resistance quantum, $L$ is the length of the wire, and $\xi(0)$ is
the zero-temperature coherence length. Equations~\eqref{R(T)} and
\eqref{F(T)} define the so-called Little's fit. Finally, coefficient
$D$ in the table was introduced for the activation energy of the PSJ
model as $U_c=D\sqrt{6}\hbar I_c/2e$.

\begin{figure}[h!]
  \includegraphics[width=7cm]{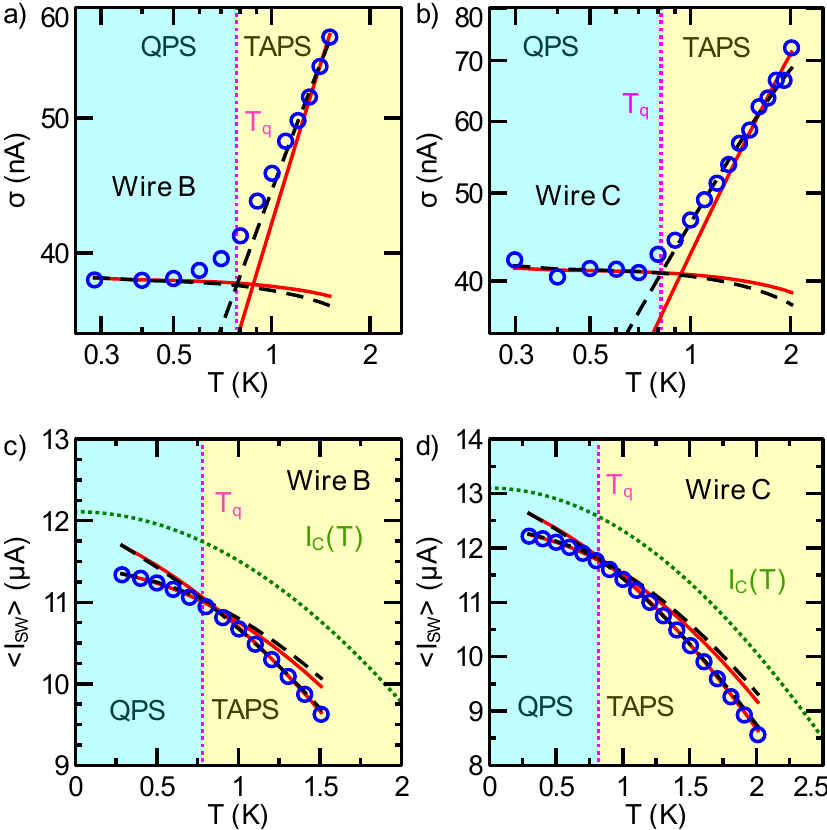}\vskip-.15cm
  \caption{[Color online] Standard deviations and critical currents versus temperature for wires
  $B$ and $C$. The convention for lines follow that explained
  in the caption of Fig.~\ref{Fig1} in the main text.}\label{SOMFig4}\vskip-.15cm
\end{figure}

For completeness, we show in Figs.~\ref{SOMFig3}-\ref{SOMFig4}
additional experimental data for the measured standard deviations
and the corresponding switching currents for the other wires listed
in the table of Fig.~\ref{Table}. These additional samples
consistently show saturation of the dispersion of the switching
current at low temperatures, where quantum phase slips proliferate.
What is of particular significance is that the saturation of the
dispersion is accompanied by the continued increase (with cooling)
of the mean switching current below the crossover temperature. The
theoretical fits are in good agreement with such observed behavior.


\begin{thebibliography}{00}
\bibitem{Gumbel}
For the rate taken from Eq.~\eqref{gamma} the analytical solution
for the current switching distribution
$P(I)=\Gamma(I)(dI/dt)^{-1}\big[1-\int^{I}_{0}P(x)dx\big]$ is in the
form of the Gumbel distribution. The Gumbel distribution is defined
by the two-parameter function
$P(I)=I_\beta^{-1}\exp\big(\frac{I-I_\alpha}{I_\beta}-e^{(I-I_\alpha)/I_\beta}\big)$.
\bibitem{Note}
A more appropriate definition of the crossover temperature between
regimes 1 and 2 could involve corresponding phase slip-rates
$\Gamma_{i}$ evaluated at a typical current which can be taken to be
the average switching current $\langle I_{sw,i} \rangle$ (we assume
``$i=1$'' is QPS and ``$i=2$'' is TAPS). Such rates characterize the
strength of phase-slip mechanisms. The crossover temperature $T_q$
is then defined by the condition $\Gamma_1(T_q, \langle I_{sw, 1}
\rangle) = \Gamma_2(T_q, \langle I_{sw, 2} \rangle)$. A drawback of
such a definition is that it involves two, in general different,
values of $\langle I_{sw,i} \rangle$, while experimentally the
switching always  occurs at some uniquely defined $\langle I_{sw}
\rangle$.
\end{thebibliography}
\end{document}